Non-equilibrium entropy and irreversibility in generalized stochastic Loewner evolution from an information-theoretic perspective

Yusuke Shibasaki[*] and Minoru Saito[†]

*Department of Correlative Study in Physics and Chemistry, Graduate School of Integrated Basic Sciences, Nihon University, Setagaya, Tokyo 156-8550, Japan*

Abstract

The generalized stochastic Loewner evolution (SLE) driven by reversible Langevin dynamics was theoretically investigated in the context of non-equilibrium statistical mechanics. The recent study of the authors revealed that the Loewner evolution enables encoding the non-equilibrium (irreversible) processes into equilibrium (reversible) processes. In this study, by Gibbs entropy-based information-theoretic approaches, we formulated this encoding mechanism of the SLE to discuss its advantages as a mean to better describe non-equilibrium states. After deriving entropy production and flux for the 2D trajectories of the generalized SLE curve, we reformulated the system's entropic properties in terms of the Kullback-Leibler (KL) divergence. We demonstrate that this operation leads to alternative expressions of the Jarzynski equality and the second law of thermodynamics, which are consistent with the previously suggested theory of information thermodynamics. The irreversibility of the 2D trajectory was likewise discussed by decomposing its entropy into additive and non-additive parts. We numerically verified the non-equilibrium property of our model by simulating the long-time behavior of the entropic measure suggested by our formulation, referred to as the relative Loewner entropy.

[*] JSPS Research Fellow: yshib@chs.nihon-u.ac.jp
[†] msaito@chs.nihon-u.ac.jp

# 1. INTRODUCTION

Although the irreversibility of non-equilibrium systems has been discussed in numerous fields for decades, the difficulties accompanying their theoretical formulation essentially involve the definition of the concept of *entropy* [1-3]. Since the pioneering study by Prigogine, et al. [4], entropy production describing the dissipative open systems far from equilibrium has been studied by employing Gibbs entropy-based approaches [5-14]. These formulations assume that non-equilibrium states are characterized by a non-zero entropy production rate of the system, and time reversibility (or time symmetry) holds only when the system is in an equilibrium state with zero entropy production rate [7-9]. These assumptions are shown by various types of the fluctuation theorem (FT) [15-17] combined with the stochastic dynamics described by Langevin and Fokker-Planck equations [8, 9, 18-20], etc. One of the advantages of the Gibbs-entropy based approach is that it is compatible with the Shannonian *information entropy*. Whereas the information entropy is a measure of uncertainty of the events consistently used for describing equilibrium states, the concept of *information* is often adopted into the theory of thermodynamics as a quantity we obtain by the measurement of the system [2, 21, 22]. Consequently, the Gibbs-Shannon entropy-based approaches for non-equilibrium systems have also been proposed in several studies [23, 24]. However, even with the above-mentioned approaches, the characterization of non-equilibrium states using quantities from equilibrium physics includes several controversial issues (e.g., the measure describing non-equilibrium stationary states). For this reason, an alternative form of the Gibbs-Shannon entropy was proposed based on the non-additivity of the ensembles of non-equilibrium states [25, 26]. These problems concerning *non-equilibrium entropy* can be reduced to the estrangement between non-equilibrium physics and well-established equilibrium physics.

A previous study of the authors indicated that the stochastic Loewner evolution (SLE) proposed by Schramm [27, 28] provides a bridge between the equilibrium and non-equilibrium (i.e., reversible and irreversible) statistical mechanics systems [29]. The SLE theory typically describes the conformally invariant geometries (curves) in various 2D statistical mechanics models, whose time evolutions are determined by the Loewner equation [30] driven by Brownian motion (Wiener process). In Ref. 29, the authors reported that the framework of the SLE can be regarded as a system that encodes the thermodynamically irreversible trajectories of the curves into the reversible driving functions. This shows, in other words, that the 2D non-equilibrium trajectories described by the SLE are the images of the equilibrium systems under the conformal transformations uniquely determined by the Loewner equation.

In this study, we develop this perspective by using a generalized SLE framework, employing a driving function governed by the Langevin equation (described in Sec. II). We present an information-theoretic perspective for the thermodynamics of the SLE to show the advantages of *encoding nonequilibrium processes into equilibrium processes*. Motivated by the above, our aim is to formulate the Gibbs-Shannon entropy-based relations between curves in the physical plane and driving functions in the mathematical plane in a generalized SLE framework (presented in Sec. III). The main tools of the first step of our analysis are the Langevin and Fokker-Planck equations describing the trajectories of the tip of the curve, which are available only when the corresponding driving function satisfies the time symmetric property [31-34]. After deriving several basic non-equilibrium entropic relations (e.g., entropy production and flux, Jarzynski equality [35]), we deduce these relations in terms of the Kullback-Leibler (KL) divergence [36-39] to introduce an extended second law of the thermodynamics. Subsequently, by considering the phase space deformation induced by the conformal maps determined by the Loewner equation, we suggest a novel irreversibility measure, which we call the *relative Loewner entropy*. We demonstrate that the relative Loewner entropy, defined as a probabilistic divergence between the trajectory of the curve and the driving function, is closely related to the Lyapunov exponent of the conformal map in the Loewner equation. Using this quantity, numerical simulations were performed to verify non-equilibrium states of the generalized SLE curves (in Sec. IV). In the discussion (Sec. V), we reinterpret the statistical physical meanings of our results, most of which rely on information theory, in relation to the problem of the determination of a non-equilibrium entropy.

## II. MODEL

### A. Chordal Loewner evolution

We consider the chordal Loewner evolution described as follows. Let $\gamma_{[0,t]}$ be a simple curve parametrized by time $t$ on the upper half complex plane $\mathbb{H}$, starting from the origin. The following Loewner equation yields a family of time-dependent conformal maps $g_t$ from $\mathbb{H} \setminus \gamma_{[0,t]}$ to $\mathbb{H}$:

$$\frac{\partial g_t(z)}{\partial t} = \frac{2}{g_t(z) - (\xi_t - \xi_0)}, \quad g_0(z) = z, \quad z \in \mathbb{H}, \tag{1}$$

where $\xi_t$ is a one-dimensional real-valued time function called the driving function [27, 28, 30] defined in the following subsection. $\xi_0$ is the initial condition of the driving function. When $\xi_t$ corresponds to the Brownian motion (i.e., Wiener process), Eq. (1) describes the SLE process in a usual sense. The conformal map satisfying the Loewner equation in Eq. (1) is given as follows [30]:

$$g_t(z) = z + \frac{2t}{z} + O(|z^{-2}|). \quad \text{as } z \to \infty. \tag{2}$$

The relation between the tip of the curve $\gamma_t$ and the driving function $\xi_t$ is expressed as follows [30]:

$$\lim_{z \to \gamma_t} g_t(z) = \xi_t - \xi_0. \tag{3}$$

Therefore, from Eqs. (1) and (2), it is evident that the family of $g_t$ encodes the history of the time evolution of the tip $\gamma_{[0,t]}$ into the driving function $\xi_{[0,t]}$. Notably, this transformation has a one-to-one correspondence between the curves and driving functions and is reversible in the sense of the uniqueness of the inverse transformation, i.e., $\lim_{w \to \xi_t - \xi_0} g_t^{-1}(w) = \gamma_t$, $(w \in \mathbb{H})$. However, in practice, the determination of $g_t$ is difficult because $g_t$ changes along continuous time. Indeed, this encoding mechanism is a physically non-trivial and meaningful process, as we show later.

B. Langevin dynamics as a driving function

We consider that the driving function $\xi_t$ of the Loewner evolution is governed by the following Langevin equation [40]:

$$\frac{d\xi_t}{dt} = \alpha(\xi_t) + \sqrt{\kappa}\frac{dB_t}{dt}, \qquad (4)$$

where $B_t$ is the Wiener process, and $\kappa$ is a diffusion parameter. The drift term $\alpha(\xi_t)$ is assumed to be a conservative force that has a potential function $V(\xi,t)$ satisfying

$$\alpha(\xi_t) = -\frac{\partial V(\xi,t)}{\partial \xi}. \qquad (5)$$

This condition guarantees that $\xi_t$ can be a time-reversible process [41]. The associated Fokker-Planck equation is described as [7]

$$\frac{\partial p(\xi,t)}{\partial t} = \left[\frac{\kappa}{2}\frac{\partial^2}{\partial \xi^2} - \frac{\partial}{\partial \xi}\alpha(\xi_t)\right]p(\xi,t). \qquad (6)$$

Here, $p(\xi,t) = \langle \delta(\xi - \xi_t) \rangle$, where the brackets denote the ensemble average. For convenience, we define the probability current as

$$J(\xi,t) = \left[\frac{\kappa}{2}\frac{\partial}{\partial \xi} - \alpha(\xi_t)\right]p(\xi,t). \qquad (7)$$

Then the Fokker-Planck equation in Eq. (6) is expressed as

$$\frac{\partial p(\xi,t)}{\partial t} = \frac{\partial}{\partial \xi}J(\xi,t). \qquad (8)$$

We assume the scenario where $\xi_t$ is in the stationary (equilibrium) state by:

$$p(\xi,t) = p_s(\xi_t) = \frac{1}{Z}e^{-2V(\xi_t)/\kappa}. \qquad (9)$$

Here, $Z$ is a normalization constant, and $p_s(\xi_t)$ is a stationary probability distribution that satisfies $J(\xi,t) = 0$ [7-9].

## III. GENERAL FORMULATIONS

### A. Equilibrium condition on mathematical plane

In the following formulations, we impose the equilibrium condition on the driving function in the mathematical plane from the initial condition, which is characterized by the zero-entropy production rate of $\xi_t$ constructed above. Let us define the Gibbs entropy of $\xi_t$ as follows:

$$S^m = - \int p_s(\xi, t) \ln p_s(\xi, t) \, d\xi = \langle s^m \rangle_{\xi_t}, \quad (10)$$

where $s^m = -\ln p_s(\xi_t)$ and $\langle \cdot \rangle_{\xi_t}$ denotes the ensemble average over all realizations of the driving function. $S^m > 0$ and $dS^m/dt = 0$ can be derived from Eqs. (9) and (10), and it indicates the non-negative and time-independent properties of the equilibrium entropy.

Furthermore, we assume the following detailed balance condition for the driving function [40],

$$p(\xi_n|\xi_{n-1})p(\xi_{n-1}|\xi_{n-2}) \cdots p(\xi_1|\xi_0)p_s(\xi_0) = p(\xi_0|\xi_1)p(\xi_1|\xi_2) \cdots p(\xi_{n-1}|\xi_n)p_s(\xi_n). \quad (11)$$

Here, $p(A|B)$ is the transition probability from state B to A and $n\,(\geq 1)$ is the integer index satisfying $t = n\tau$ where $\tau$ is a sufficiently small-time interval. Let us define

$$P[\xi_{path}(n)] = p(\xi_n|\xi_{n-1})p(\xi_{n-1}|\xi_{n-2}) \cdots p(\xi_1|\xi_0)p_s(\xi_0), \quad (12)$$

and

$$\tilde{P}[\tilde{\xi}_{path}(n)] = p(\xi_0|\xi_1)p(\xi_1|\xi_2) \cdots p(\xi_{n-1}|\xi_n)p_s(\xi_n). \quad (13)$$

Then, we define the ratio between these probability as $R_m = P[\xi_{path}(n)]/\tilde{P}[\tilde{\xi}_{path}(n)]$, so that

$$\ln R_m \equiv \ln \frac{P[\xi_{path}(n)]}{\tilde{P}[\tilde{\xi}_{path}(n)]} = 0, \quad (14)$$

which follows from Eq. (11). From the formulation using the master equations [7-10, 42], $\ln R_m = 0$ suggests that there is no entropy production inside the system for each trajectory and the microscopic time reversibility is guaranteed for all time.

B. Entropy production in physical plane

We investigate the entropy production of the trajectory of the curve in the physical plane. We demonstrate the irreversible and dissipative character of the SLE curve, which differs from that of the driving function. We mainly use the Langevin and Fokker-Planck equation for them, which are available for the detailed balanced condition. The formulation using the backward Loewner evolution [31-34] shows that if the driving function is time-symmetric (i.e., $-\xi_t$ and $\xi_{-t}$ have the same probability distribution) and has stationary increments, the probability distribution for the time evolution of the tip of the curve $z_t$ is the same of that of $(x_t, y_t)$, described by the following two-dimensional Langevin equation.

$$\frac{dx_t}{dt} = -\frac{2x_t}{x_t^2 + y_t^2} - \alpha(\xi_t) - \sqrt{\kappa}\frac{dB_t}{dt}$$

$$\frac{dy_t}{dt} = \frac{2y_t}{x_t^2 + y_t^2}. \tag{15}$$

Here, we adopt the initial condition of $x_0 = 0$ and $y_0 = \varepsilon$, where $\varepsilon$ is an infinitesimal positive constant [32]. The Fokker-Planck equation associated with Eq. (15) is expressed as follows.

$$\frac{\partial p(x,y,t)}{\partial t} = \left\{\frac{\kappa}{2}\frac{\partial^2}{\partial x^2} + \frac{\partial}{\partial x}\left[\frac{2x}{x^2+y^2} + \alpha(\xi_t)\right] - \frac{\partial}{\partial y}\frac{2y}{x^2+y^2}\right\}p(x,y,t). \tag{16}$$

Here, $p(x, y, t) = \langle \delta(x - x_t)\delta(y - y_t)\rangle$, where the ensemble average is calculated over all realizations of the curves. For the latter formulations, we take

$$\chi(x,y,t) = \frac{\kappa}{2}\frac{\partial}{\partial x}p(x,y,t) + \left[\frac{2x}{x^2+y^2} + \alpha(\xi_t)\right]p(x,y,t) \tag{17}$$

and

$$\psi(x,y,t) = \frac{2x}{x^2+y^2}p(x,y,t). \tag{18}$$

Substituting Eqs. (17) and (18) into Eq. (16), the Fokker-Planck equation for the trajectory of curve is expressed as

$$\frac{\partial p(x,y,t)}{\partial t} = \frac{\partial}{\partial x}\chi(x,y,t) - \frac{\partial}{\partial y}\psi(x,y,t). \tag{19}$$

Subsequently, we define the following time-dependent Gibbs entropy for the trajectory of the curve.

$$S^p(t) = -\iint p(x,y,t)\ln p(x,y,t)\,dxdy = \langle s^p \rangle_{x_t,y_t}, \tag{20}$$

where $s^p = -\ln p(x,y,t)$, and $\langle \cdot \rangle_{x_t,y_t}$ denotes the ensemble average over all possible realizations of $(x_t, y_t)$. We are interested in the changing rate of $S^p(t)$, which are formulated by Prigogine et al. in the following [4, 7, 8, 29]:

$$\frac{dS^p(t)}{dt} = \frac{dS_i^p(t)}{dt} - \frac{dS_e^p(t)}{dt}, \tag{21}$$

where $dS_i^p(t)/dt$ is the entropy production rate inside the system, which is non-negative because of the second law of thermodynamics. The second term in the right-hand side (r.h.s.) of Eq. (21), $dS_e^p(t)/dt$ is the entropy flux rate from the system to the external environment. If the system is stationary, $dS_i^p(t)/dt = dS_e^p(t)/dt$, whereas if the system is in equilibrium, $dS_i^p(t)/dt = dS_e^p(t)/dt = 0$. In both scenarios, $S^p(t)$ assumes a constant value, otherwise $S^p(t)$ changes depending on the time and the system is in non-equilibrium [7, 8]. For subsequent discussions, we define the entropy production $s_i^p$ and entropy flux $s_e^p$ for the individual trajectories of the tips of the curves as those satisfying $S_i^p(t) = \langle s_i^p \rangle_{x_t,y_t}$ and $S_e^p(t) = \langle s_e^p \rangle_{x_t,y_t}$.

Hereafter, we apply the entropic formulation in Eq. (21) to the SLE curve in our model, using Eqs. (17)-(20), and performing partial integrations; the time derivative of $S^p(t)$ can be calculated as [29]

$$\frac{dS^p}{dt} = \iint \frac{2}{\kappa}\frac{[\chi(x,y,t)]^2}{p(x,y,t)}dxdy - \iint \frac{2}{\kappa}\left(\frac{2x}{x^2+y^2} + \alpha(\xi_t)\right)\chi(x,y,t)\,dxdy$$
$$- \iint \frac{2y}{x^2+y^2}\frac{\partial}{\partial v}p(x,y,t)dxdy. \tag{22}$$

Here, we dropped the boundary terms whose $p(x,y,t)$ tends to zero when $x \to \pm\infty$ or $y \to \pm\infty$. Because the first term of the r.h.s. of Eq. (22) is non-negative, we can identify it with the entropy production rate, that is:

$$\frac{dS_i^p(t)}{dt} = \iint \frac{2}{\kappa} \frac{[\chi(x,y,t)]^2}{p(x,y,t)} dxdy \geq 0. \tag{23}$$

The equality holds when $\chi(x,y,t) = 0$, and this is a necessary condition for thermal equilibrium. In this framework, the entropy production rate is described in terms of the free energy $F$ of the system as $dS_i^p(t)/dt = -dF/dt$ [13]. Therefore, by combination with Eq. (23), $dF/dt \leq 0$ can be derived. This outcome is interpreted as the *H*-theorem for the trajectory of the SLE curve, and it ensures that the system is thermodynamically irreversible in time, except for the equilibrium condition [43].

Meanwhile, the second and third terms of r.h.s. of Eq. (22) are interpreted as the contributions for the entropy flux rate. Performing the partial integration and using the definition of the ensemble average, the entropy flux rate is expressed as [29]:

$$\frac{dS_e^p(t)}{dt} = \left\langle \frac{2}{\kappa}\left(\frac{2x_t}{x_t^2 + y_t^2} + \alpha(\xi_t)\right)^2 - \frac{\partial \alpha(\xi_t)}{\partial x} \right\rangle_{x_t, y_t}. \tag{24}$$

Then, the entropy flux for individual trajectory $s_e^p$ (from $t=0$) is calculated by the following integral:

$$s_e^p = \int_0^t \left[\frac{2}{\kappa}\left(\frac{2x_t}{x_t^2 + y_t^2} + \alpha(\xi_t)\right)^2 - \frac{\partial \alpha(\xi_t)}{\partial x}\right] dt, \tag{25}$$

which is the total amount of entropy dissipated to the external environment for each trajectory of the curves. Note that $(x_t, y_t)$ and $(\text{Re}(z_t), \text{Im}(z_t))$ have the same joint probability distribution. The increase of $s_e^p$ indicates that the generalized SLE curve remains an open non-equilibrium process, contrary to the fact that the corresponding driving function is an equilibrium process from the initial states.

### C. Jarzynski equality for generalized SLE curve

We derive the equality governing the time irreversible trajectories of the SLE curve, which was originally derived by Jarzynski [35] and applied to stochastic trajectories by Seifert [10]. Let us denote the discretized points on curve $\gamma_{[0,t]}$ as $\gamma_{[0,n]} = \{z_0(=0), z_1, z_2, ..., z_n\}$. Then, in the same manner as Eqs. (12) and (13), we define

$$P[z_{path}(n)] = p(z_n|z_{n-1})p(z_{n-1}|z_{n-2})\cdots p(z_1|z_0)p(z_0) \qquad (26)$$

and

$$\tilde{P}[\tilde{z}_{path}(n)] = p(z_0|z_1)p(z_1|z_2)\cdots p(z_{n-1}|z_n)p(z_n). \qquad (27)$$

Subsequently, we define the ratio of these probabilities as $R_p = P[z_{path}(n)]/\tilde{P}[\tilde{z}_{path}(n)]$. From the formulations using master equations, $\ln R_p$ is expressed in terms of the entropy flux as follows [8, 10]:

$$\ln R_p \equiv \ln \frac{P[z_{path}(n)]}{\tilde{P}[\tilde{z}_{path}(n)]} = s_e^p + \ln \frac{p(z_0)}{p(z_n)}. \qquad (28)$$

Here, the individual entropy flux $s_e^p$ is given by Eq. (25). From Eq. (28), Jarzynski equality can be derived as [8, 10]

$$\langle e^{-\ln R_p}\rangle_{z_{path}} = \sum_{z_{path}} P[z_{path}(n)] e^{-\ln R_p}$$

$$= \sum_{\tilde{z}_{path}} \tilde{P}[\tilde{z}_{path}(n)]$$

$$= 1. \qquad (29)$$

Here, $\langle\cdot\rangle_{z_{path}}$ denotes the ensemble average over the forward path of the curve. Using Eqs. (25) and (28), Eq. (29) is expressed as

$$\left\langle \exp\left\{-\int_0^t \left[\frac{2}{\kappa}\left(\frac{2x_t}{x_t^2 + y_t^2} + \alpha(\xi_t)\right)^2 - \frac{\partial \alpha(\xi_t)}{\partial x}\right] dt + \ln \frac{p(z_n)}{p(z_0)}\right\}\right\rangle_{z_{path}} = 1. \qquad (30)$$

This is the Jarzynski equality for the generalized SLE curve, which is applicable regardless of whether system is in equilibrium or nonequilibrium.

### D. KL divergence for generalized SLE

Although we showed the underlying entropic law in the trajectories of generalized SLE curves using previously studied formulations, the nonequilibrium characteristic of the 2D trajectory in the physical plane is generated by the transformation of the reversible driving function. To clarify this encoding property of the SLE in terms of information theory, we take an approach using the KL divergence. In the following formulation, we eliminate the restrictions on initial conditions, assuming the that the driving function is in a relaxation process to the equilibrium state, which requires a certain length of time.

Subtracting Eq. (28) from Eq. (14) yields

$$\ln(R_m/R_p) = \ln\frac{P[\xi_{path}(n)]}{\tilde{P}[\tilde{\xi}_{path}(n)]} - \ln\frac{P[z_{path}(n)]}{\tilde{P}[\tilde{z}_{path}(n)]}$$
$$= \ln\frac{\tilde{P}[\tilde{z}_{path}(n)]}{\tilde{P}[\tilde{\xi}_{path}(n)]} - \ln\frac{P[z_{path}(n)]}{P[\xi_{path}(n)]}. \quad (31)$$

We define KL divergences between the forward paths of the driving function and the curve as the following:

$$D(z_{path}\|\xi_{path}) = \sum_{z_{path}(n)} P[z_{path}(n)]\ln\frac{P[z_{path}(n)]}{P[\xi_{path}(n)]}. \quad (32)$$

Similarly, for the backward paths we define

$$\tilde{D}(\tilde{z}_{path}\|\tilde{\xi}_{path}) = \sum_{\tilde{z}_{path}(n)} \tilde{P}[\tilde{z}_{path}(n)]\ln\frac{\tilde{P}[\tilde{z}_{path}(n)]}{\tilde{P}[\tilde{\xi}_{path}(n)]}. \quad (33)$$

Then, we denote $d(z_{path}\|\xi_{path}) = \ln\frac{P[z_{path}(n)]}{P[\xi_{path}(n)]}$ and $\tilde{d}(\tilde{z}_{path}\|\tilde{\xi}_{path}) = \ln\frac{\tilde{P}[\tilde{z}_{path}(n)]}{\tilde{P}[\tilde{\xi}_{path}(n)]}$, so that

$$D(z_{path}\|\xi_{path}) = \langle d(z_{path}\|\xi_{path})\rangle_{z_{path}},$$

and

$$\tilde{D}(\tilde{z}_{path}\|\tilde{\xi}_{path}) = \langle \tilde{d}(\tilde{z}_{path}\|\tilde{\xi}_{path})\rangle_{\tilde{z}_{path}}. \quad (34)$$

Using these expressions, Eq. (31) is expressed as

$$R_m/R_p = \exp[\tilde{d}(\tilde{z}_{path}\|\tilde{\xi}_{path}) - d(z_{path}\|\xi_{path})]. \quad (35)$$

For the left-hand side of Eq. (35), we take the dual ensemble average with respect to $z_{path,}$ and $\tilde{\xi}_{path}$,

hence,

$$\langle R_m/R_p \rangle_{z_{path}, \tilde{\xi}_{path}} = \sum_{\tilde{\xi}_{path}} \sum_{z_{path}} P[z_{path}(n)] \tilde{P}[\tilde{\xi}_{path}(n)] e^{\ln(R_m/R_p)}$$

$$= \sum_{\xi_{path}} \sum_{\tilde{z}_{path}} \tilde{P}[\tilde{z}_{path}(n)] P[\xi_{path}(n)]$$

$$= 1. \tag{36}$$

Because we obtain the relation $\langle R_m/R_p \rangle_{z_{path}, \tilde{\xi}_{path}} = 1$, using Eq. (36), Eq. (35) yields

$$\langle \exp[\tilde{d}(\tilde{z}_{path}\|\tilde{\xi}_{path}) - d(z_{path}\|\xi_{path})] \rangle_{z_{path}, \tilde{\xi}_{path}} = 1. \tag{37}$$

This is an alternative of the Jarzynski equality, which is applicable for the generalized SLE that we defined previously. From the Jensen's inequality, the following relation is derived:

$$\langle \exp[\tilde{d}(\tilde{z}_{path}\|\tilde{\xi}_{path}) - d(z_{path}\|\xi_{path})] \rangle_{z_{path}, \tilde{\xi}_{path}}$$

$$\geq \exp\left[\langle \tilde{d}(\tilde{z}_{path}\|\tilde{\xi}_{path}) \rangle_{z_{path}, \tilde{\xi}_{path}} - \langle d(z_{path}\|\xi_{path}) \rangle_{z_{path}, \tilde{\xi}_{path}}\right]. \tag{38}$$

Substituting Eq. (37) into Ineq. (38), we obtain the following inequality,

$$\langle D(z_{path}\|\xi_{path}) \rangle_{\tilde{\xi}_{path}} \geq \langle d(z_{path}\|\xi_{path}) - \ln R_p + \ln R_m \rangle_{z_{path}, \tilde{\xi}_{path}}. \tag{39}$$

Using Eq. (28), Ineq. (39) is transformed as follows:

$$0 \leq \left\langle s_e^p + \ln\frac{p(z_0)}{p(z_n)} - \ln R_m \right\rangle_{z_{path}, \tilde{\xi}_{path}} \tag{40}$$

Using the relation $s_i^p = s_e^p + \ln\frac{p(z_0)}{p(z_n)}$, the following inequality is derived,

$$0 \leq \langle s_i^p - \ln R_m \rangle_{z_{path}, \tilde{\xi}_{path}}, \tag{41}$$

This relation can be interpreted as an extension of the second law of the thermodynamics. In Ineq. (41), the equality holds true if the trajectory is in an equilibrium state characterized by $s_i^p = \ln R_m = 0$. Note that $s_e^p$ in Eq. (40) is expressed by Eq. (25) only when the time reversibility of the driving function is guaranteed.

E. Relative Loewner entropy

We showed that the non-equilibrium states of the trajectory of the SLE curves are formulated by the Shannon entropy-based KL divergence. Furthermore, we demonstrated the unique property of the Loewner evolution by characterizing the entropy flux production and flux terms in terms of the path probabilities of the curves and the driving function. In this subsection, we elucidate the intrinsic mechanism that affords this encoding mechanism by examining the phase volume deformation induced by the conformal map $g_t$, which is shown to be a fundamental factor of the irreversibility. We further incorporate a viewpoint of the non-additivity of the entropy, which is a basic concept of non-extensive statistical mechanics [25, 26]. In the followings, we assume that the driving function is in an equilibrium state.

The relation between the probability for $z$ and $w$ under the conformal map $z = g_t^{-1}(w)$ is expressed as [44, 45]

$$p(z) = p(w) \left| \frac{dg_t(z)}{dz} \right|^2. \tag{42}$$

Using the relation $\xi_t - \xi_0 = \lim_{z \to \gamma_t} g_t(z)$ $(\gamma_t = z_t)$ in Eq. (3) and taking the logarithms, the entropy $s^p(z)$ for the tip on the curve is described as

$$s^p(z) = s^m(\xi) - \ln \left| \frac{dg_t(z_t)}{dz} \right|^2, \tag{43}$$

where from Eq. (2) for large $z_t$,

$$\left| \frac{dg_t(z_t)}{dz} \right| = \left| 1 - \frac{2t}{z_t^2} + O(|z_t^{-3}|) \right|. \tag{44}$$

Here, the entropy $s^p(z)$ is decomposed into an equilibrium entropy $s^m(\xi)$, and the rest of the part $-\ln \left| \frac{dg_t(z_t)}{dz} \right|^2$. Because $s^m(\xi)$ is associated with the time-independent canonical distribution $p_s(\xi_t)$ described by Eq. (9), it can be referred to as the *additive* part. Contrarily, $\ln \left| \frac{dg_t(z_t)}{dz} \right|^2$ is time-dependent

as we will numerically show later, and referred to as the *non-additive* part in the sense of the non-extensive statistical mechanics [25, 26, 46]. Notably, the behavior of the non-additive part characterizes non-equilibrium (irreversible) properties of the generalized SLE curve. Let us denote $d(z_t \parallel \xi_t) = \ln \left| \frac{dg_t(z_t)}{dz} \right|^2$. From Eq. (43), the time-derivative of $s^p(z)$ is expressed as:

$$\frac{ds^p(z)}{dt} = -\frac{d}{dt} d(z_t \parallel \xi_t). \tag{45}$$

Here, we used $\partial p_s(\xi, t)/\partial t = 0$ from $J = 0$. In Eq. (45), non-zero of $ds^p(z)/dt$ indicates the time-dependence of $s^p(z)$, and the non-equilibrium states of the individual curve trajectory. Furthermore, $d(z_t \parallel \xi_t) \to 0$ indicates the relaxation to thermal equilibrium states, and $d(z_t \parallel \xi_t) \to const. (\neq 0)$ indicates the convergence to other (non-equilibrium) stationary states.

Taking the ensemble average with respect to $z_t$, the KL divergence between $z_t$ and $\xi_t$ is derived as:

$$D(z_t \parallel \xi_t) = \langle d(z_t \parallel \xi_t) \rangle_{z_t} \simeq \left\langle \ln \left| 1 - \frac{2t}{z_t^2} + O(|z_t^{-3}|) \right|^2 \right\rangle_{z_t}, \tag{46}$$

This quantity works as an indicator of the irreversibility and stationarity of the whole ensemble of the trajectories of the curves. We call $D(z_t \parallel \xi_t)$ [or $d(z_t \parallel \xi_t)$ depending on the situations] expressed by Eq. (46) the *relative Loewner entropy*, which is used to estimate the nonequilibrium state of the 2D trajectory on $\mathbb{H}$.

The relative Loewner entropy has a connection with phase volume deformation under the conformal map $g_t$. If we take the time average of the non-additive part in Eq. (43) after calculus, we obtain a form of the Lyapunov exponent [47] for the conformal map $g_t$,

$$\lambda = \lim_{T \to \infty} \frac{1}{T} \sum_{t=0}^{T} \ln \left| \frac{dg_t(z_t)}{dz} \right| = \frac{1}{2} \overline{d(z_t \parallel \xi_t)}, \quad (T \gg 0). \tag{47}$$

Note that the overbar represents the time average. Here, the Lyapunov exponent $\lambda$ defined by Eq. (47) measures, rather than the sensitivity to the initial condition, the time averaged phase volume expansion ($\lambda > 0$) and contraction ($\lambda < 0$) of the neighborhood of the tip of the curve on $\mathbb{H}$ under the map $g_t(z_t)$. Equations (45) and (47) show that it closely related to the total entropy production rate of $S^p(z)$. Note, that an equilibrium state is characterized by $\lambda = 0$ from Eq. (47).

### IV. NUMERICAL TESTS

To realize the curves $\gamma_{[0,t]}$ of our model, numerical simulations were performed in the following methods. First, Langevin dynamics in Eq. (4) was simulated by choosing the potential function as $V(\xi) = \frac{1}{2}a\xi^2$, such that $\alpha(\xi_t) = -a\xi_t$, where $a$ is a positive constant. Consequently, the driving function can be described by a linear Langevin equation. The discretization of the Langevin equation is performed in the similar method in Ref. 48, that is,

$$\xi_i = \xi_{i-1} - a\tau\xi_{i-1} + \sqrt{\kappa\tau}W_{i-1}, \quad (i \geq 1) \tag{48}$$

Here, $\tau$ is a sufficiently small unit time interval, and $W_i$ is the white Gaussian noise with mean 0 and variance 1.0. $\kappa$ is the diffusion parameter which is the same as that in Eq. (4). The initial condition is set as $\xi_0 = \sqrt{\kappa/2a}$, which is the condition derived from fluctuation dissipation theorem [49], that the driving function is in an equilibrium state from the initial state. After simulating Langevin dynamics using Eq. (48), the shifted driving function $\xi_i - \xi_0$ was calculated such that its initial condition is zero. We note that this operation makes the curves start at origin in the theoretical scheme, however, this is not necessary for our numerical computation algorithm described below, because we use the time differences of $\xi_i$ only. For the numerical realizations of the curves $\gamma_{[0,n]}$, we employed the zipper algorithm using the map derived from the vertical slit map [50, 51], which is described as follows.

$$\gamma_{[0,n]} = \{z_0(=0), z_1 = f_1(0), z_2 = f_1 \circ f_2(0), \ldots, z_n = f_1 \circ \cdots \circ f_{n-1} \circ f_n(0)\}, \tag{49}$$

where

$$f_i(z) = \Delta\xi_i + \sqrt{z^2 - 4\tau}, \quad \Delta\xi_i \equiv \xi_i - \xi_{i-1}. \tag{50}$$

Figure 1 shows the examples of the curves $\gamma_{[0,n]}$ calculated using the above algorithm ($n = 1.0 \times 10^5$ and $\tau = 1.0 \times 10^{-4}$.) The drift term was chosen as $a = 1.5$. [Fig. 1(a) for $\kappa = 2.0$, 1(b) for $\kappa = 4.0$, 1(c) for $\kappa = 6.0$, and 1(d) for $\kappa = 8.0$.]

The numerical experiments using the relative Loewner entropy are aimed towards verifying the non-stationary properties of the individual trajectories of the tip $z_t$ on $\mathbb{H}$ calculated by the above procedures. Particularly, we estimate $\kappa$- and $a$- dependences to their dynamical regimes. Figure 2(a)-(c) show the temporal behaviors of $d(z_t \| \xi_t)$, calculated by $\ln\left|1 - \frac{2t}{z_t^2}\right|^2$, $a = 0.5$, $1.0$, and $1.5$, respectively. Each figure includes the plots for $\kappa = 2.0, 4.0, 6.0,$ and $8.0$. For $a = 0.5$, the $d(z_t \| \xi_t)$ fluctuated violently particularly for large $\kappa$, even after a long time passed [Fig. 2(a)]. For $a = 1.0$, there were less violent fluctuations in $d(z_t \| \xi_t)$ than those for $a = 0.5$, and they seemed to loosely converge to positive values except for $\kappa = 8.0$ [Fig. 2(b)]. For $a = 1.5$, the convergence of $d(z_t \| \xi_t)$ was more valid than that for $a = 1.0$, which indicates the nonequilibrium stationary state of the trajectory of $z_t$ [Fig. 2(c)]. These results indicate the tendency that the large $\kappa$ and the small $a$ result in the non-stationarity of the trajectories. Notably, even after the time steps $n = 1.0 \times 10^5$, the convergence of $d(z_t \| \xi_t)$ to zero, which indicates an equilibrium state, was not observed.

## V. DISCUSSIONS

We have formulated nonequilibrium statistical mechanics of a generalized SLE using an information-thermodynamical approach. The SLE framework provides a unique information-theoretic scheme, in which non-equilibrium irreversible systems (i.e., the curves in the physical plane) are encoded into reversible equilibrium systems (i.e., the driving function in the mathematical plane). We showed that this encoding operation is available owing to the one-to-one correspondence between the curves and driving function, and the phase space deformation of the conformal map $g_t$. The advantages of encoding a 2D non-equilibrium trajectory into reversible Langevin dynamics are summarized as follows.

1. The Jarzynski equality and the second law of thermodynamics were generalized in terms of information theory. The second law for our result $0 \leq \langle s_i^p - \ln R_m \rangle$ in Eq. (41) is an extension of that of Seifert's expression $0 \leq \langle s_{tot} \rangle$ (See, Refs 10 and 19). Furthermore, the term $\ln R_m$ can also be interpreted as the feedback information term, denoted as $I$ in Ref. 23, in Sagawa's information thermodynamics. Hence, incorporating the relaxation process to an equilibrium into the theory of the nonequilibrium dynamics enables us to extend the existing thermodynamical laws in the information-theoretic senses. This means that for an arbitrary 2D trajectory on $\mathbb{H}$, the validity of the second law in the usual sense $0 \leq \langle s_i^p \rangle$ is supported by the complete time reversibility of the corresponding driving function, and otherwise (i.e., if the driving function includes several irreversible characters), we must reuptake the generalized second law $0 \leq \langle s_i^p - \ln R_m \rangle$.

2. The entropy describing the non-equilibrium states of the individual trajectory is decomposed into additive and non-additive parts. This provides us with a novel non-equilibrium entropic measure, which we refer to as the relative Loewner entropy. In the sense that the non-equilibrium ensemble is decomposed into an equilibrium one and a certain function, our result in Eq. (43) is analogous to the result of Penrose, et al. [52].

3. If the driving function is in equilibrium, the relative Loewner entropy is used to determine the non-stationarity of the 2D trajectories in the physical plane. This quantity indicates the phase space deformation under the conformal map $g_t$, and it is closely related to the Lyapunov exponent.

These suggest that nonequilibrium processes are well-understood when we assume the associated equilibrium processes in the theoretical framework. The equilibrium driving function works as an idealized thermal state (or an information source), which is one of the stationary states that the curves potentially reach in the long-time limit. If the entropy of the curve trajectory completely coincides with that of driving function, in which the KL divergences become zero, the thermal equilibrium of the physical and mathematical planes can be equally characterized. Then, we encounter the problem of interpreting our information-theoretic formulations in terms of thermodynamics. However, this seems to be a difficult problem because the physical definition of the driving function is not settled yet. Let us assume that the driving function is an information (entropy) source manipulating the 2D curve trajectories, which has full information about their equilibrium states. From the results of this study, we can at least conclude that:

1. If the entropy (information) of driving function is completely communicated to the physical plane, the 2D trajectories were in the equilibrium states.

2. The non-equilibrium property of the trajectories was induced by the incomplete communication of the entropy (information) between the physical and mathematical planes.

3. The driving function can work as Maxwell's Damon in the sense that it can control the feedback information $\ln R_m$, by discarding its own entropy (information).

These statements will make better sense when the integration of the information theory and thermodynamics will be successfully done.

Most importantly, the two thermodynamically different systems in our model are linked via a family of conformal maps $g_t$, determined by the Loewner equation. Therefore, as indicated in Ref. 29, they are at least mathematically convertible. Hence, in a generalized SLE framework, the microscopically irreversible process can arise from a reversible (but time-dependent) transformation to the microscopically reversible process. If we attempt to reinterpret the Boltzmann's paradox [53] in our model, the intrinsic irreversibility is caused by the small deviation between the two entropies [as shown in Eq. (43)] induced by conformal transformations. In this regard, the relative Loewner entropy explicitly represents an exact difference between non-equilibrium and equilibrium state, therefore, it can be considered as a candidate of a non-equilibrium entropy.

Moreover, although non-equilibrium physics lacks the conserved quantities, such as the Hamiltonian in the equilibrium physics, our result indicates that the entropy (or associated energy function) of the driving function is conserved, even if the corresponding curve retains in the nonequilibrium state. Therefore, the entropy of the Loewner driving function for the 2D trajectories has an important function to aid in the understanding of the irreversible and non-equilibrium processes. A similar perspective was suggested in the mathematical context, where the importance of the *energy* of the Loewner driving function was demonstrated [54, 55].

To apply our formulations, including the suggested irreversibility measure, to other various 2D self-organization phenomena (e.g., Ising systems, percolation models, turbulence, or biological/chemical morphology, etc.), we must extend the descriptive ability of the SLE framework. In the present model, we chose the driving function as Langevin dynamics with a (linear or nonlinear) drift term. This is an example of a generalization of the SLE; however, numerous other possibilities remain for the extension of the SLE framework (e.g., combining with chaotic dynamical systems [29, 34, 56], *q*-deformation [57]). Therefore, future research following this study will clarify how our theoretical concepts aid to understand real self-organization phenomena, by considering appropriate approaches to the generalization of the SLE.

## VI. CONCLUSIONS

We formulated the generalized SLE driven by Langevin dynamics in the equilibrium state from the context of nonequilibrium statistical mechanics. The entropy production of the curve trajectories in the physical plane assumed a form of irreversible nonequilibrium systems, whereas the driving function was prepared in the reversible equilibrium system. We derived alternative types of the Jarzynski equality and the second law of thermodynamics by information-theoretic quantities. Furthermore, we showed, from the phase deformation ratio of the conformal maps, that the entropy of the curve can be decomposed into additive and non-additive parts. The non-additive part was numerically examined to estimate non-equilibrium properties of the system, and we refer to it as the relative Loewner entropy. These results suggest a novel perspective for non-equilibrium statistical physics to answer the question concerning the definition of a non-equilibrium entropy and the mechanisms of irreversibility.

**Acknowledgements**

This work was supported by JSPS KAKENHI Grant Number JP20J20867 (to Y. S.).

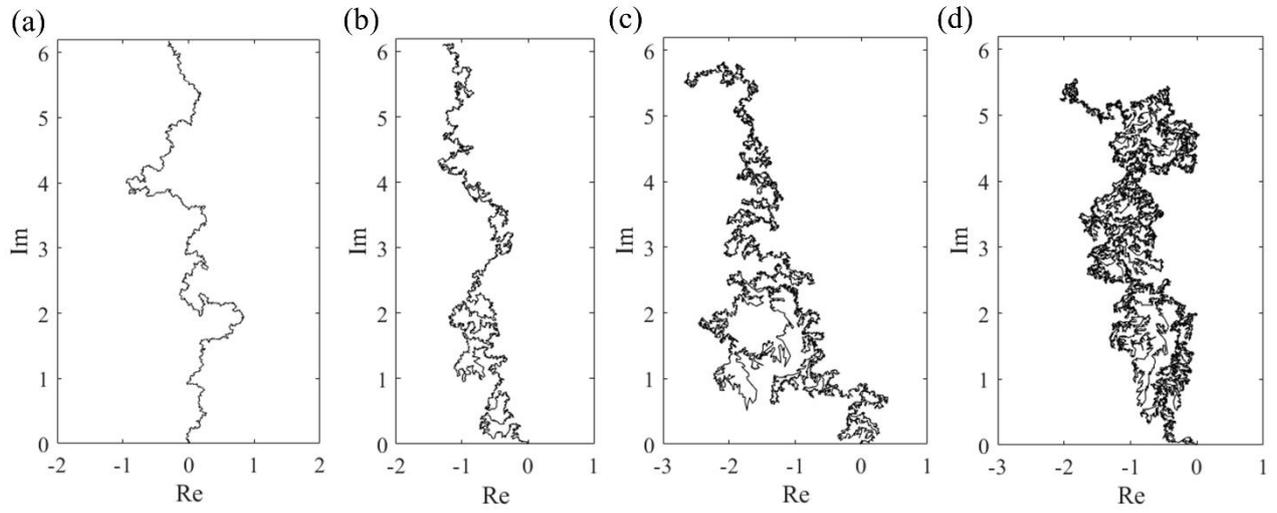

Fig.1. Typical examples of numerically realized curves $\gamma_{[0,n]}$ of generalized SLE on upper half-plane $\mathbb{H}$. Simulations were performed with $n = 1.0 \times 10^5$ and $\tau = 1.0 \times 10^{-4}$. The drift term was chosen as $a = 1.5$. (a) $\kappa = 2.0$, (b) $\kappa = 4.0$, (c) $\kappa = 6.0$, and (d) $\kappa = 8.0$.

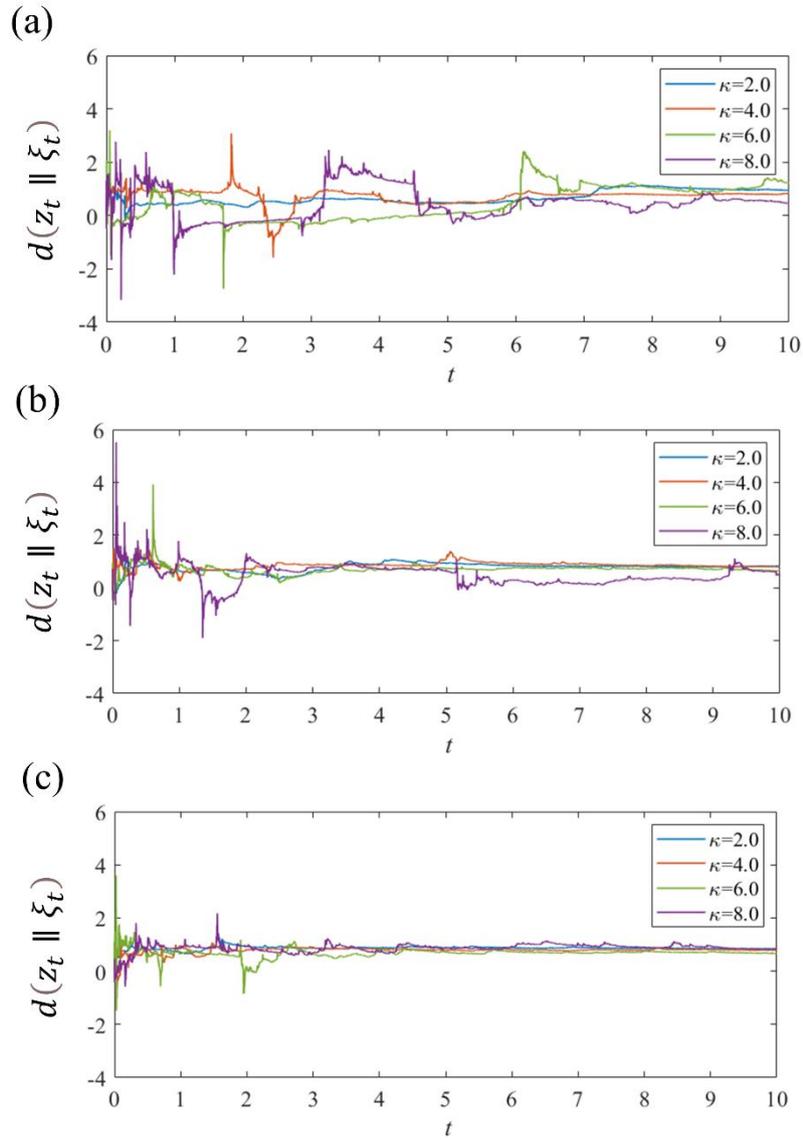

**Fig.2.** (Color) Temporal behaviors of relative Loewner entropy $d(z_t \parallel \xi_t)$ calculated as $\ln\left|1 - \frac{2t}{z_t^2}\right|^2$. (a - c) show those in the condition $a = 0.5, 1.0$ and $1.5$, respectively. Each figure includes plots for $\kappa = 2.0, 4.0, 6.0,$ and $8.0$.